\documentclass[letterpaper]{jpconf}

\usepackage{citesort}
\usepackage[dvips]{graphicx}
\usepackage{amsmath}

\begin{document}

\title{Analysis of alpha backgrounds in the DEAP-1 detector}

\author{Kevin Olsen, on behalf of the DEAP/CLEAN Collaboration}
\address{Department of Physics, University of Alberta, Edmonton T6G~2G7, Canada}
\ead{ksolsen@phys.ualberta.ca}

\begin{abstract}
  DEAP-1 is a 7 kg liquid argon dark matter detector used to prototype the tonne scale
  DEAP-3600 detector at SNOLAB.  We present an
  analysis of the alpha particle backgrounds in DEAP-1 and isolate the α radiations from various $^{222}$Rn
  daughters at various locations within the detector. The backgrounds will be removed by event position
  reconstruction and strict controls of material purity.
\end{abstract}

\section{Introduction} 
The \textbf{DEAP} (\textbf{D}ark matter \textbf{E}xperiment in \textbf{A}rgon using \textbf{P}ulse-shape 
discrimination) experiment is a direct search for spin independent scattering of WIMPs.  DEAP-3600 will 
consist of a 3600 kg (1000 kg fiducial) liquid argon target encased in acrylic and surrounded with an array of 
photo-multiplier tubes.  Located underground at SNOLAB, DEAP-3600 will achieve a very low rate of background
signals, further reduced by pulse-shape discrimination (PSD), and will allow for a WIMP-nucleon cross-section 
sensitivity of 10$^{-46}$cm$^{2}$.  These proceedings introduce the DEAP experiment and present an analysis of 
background signals due to high energy $\alpha$ particles.  We will demonstrate that while $\alpha$ radiation 
is indistinguishable from nuclear recoils using PSD alone, event position reconstruction and energy calculation
will allow us to discriminate accurately.

\begin{figure}[ht]
  \includegraphics[width=21pc]{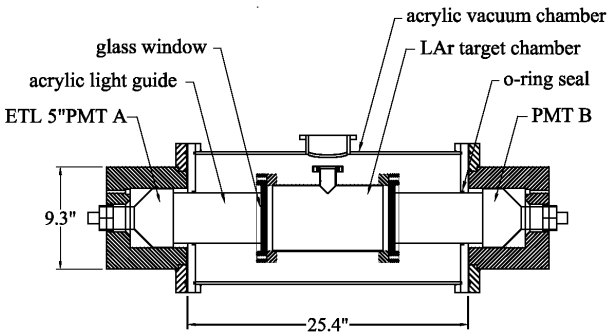}\hspace{1pc}
  \begin{minipage}[b]{15pc}\caption{\label{label}The DEAP-1 detector configuration. Liquid argon is contained in 
    a stainless steel target chamber, which includes an inner acrylic cylinder and a diffuse reflector. 
    Scintillation light from the liquid argon is wavelength-shifted by TPB on the inner detector surface 
    and is transmitted to PMTs operating at room temperature through PAMMI acrylic light guides.}
  \end{minipage}
\end{figure}

DEAP is being designed and built in two phases;  DEAP-3600,
is under development while a smaller prototype detector, DEAP-1, is taking data underground at SNOLAB
(see figure 1).  When high energy particles enter the target volume they will deposit energy in the liquid argon.  Argon 
atoms in either an excited or ionized state form excimer pairs which produce scintillation light when they dissociate\cite{KubotaDoke78}.
There are two distinct states, a triplet and singlet with lifetimes of about 1.6 $\mu s$ and 6 ns, 
respectively\cite{HitachiDoke83,Kubota82}.  The ratio of the number of singlet and triplet excimer states produced depends on 
linear energy transfer (LET).  Different incident particles, with different LETs, produce distinct signals in 
the DEAP-1 detector\cite{Lippincott08}, with nuclear recoils mainly populating the singlet state.  
This allows us to separate $\beta$-$\gamma$ interactions from nuclear 
recoils, neutrons, and $\alpha$ radiation using pulse-shape discrimination\cite{BoulayHime06,MGB09}.  The DEAP
collaboration will look for WIMPs scattering off argon nuclei, so understanding and being able to  distinguish
among other events is key.

In November 2008 the argon in DEAP-1 was topped up to maintain the maximum target volume, 
and a burst of $alpha$ activity from $^{222}$Rn was introduced into DEAP-1.
$^{222}$Rn contamination comes from argon, as they are present in the argon gas's source and 
transportation materials.
The data collected gave us unique circumstances to attempt to distinguish the energy spectra of 
$\alpha$ radiation from short lived $^{222}$Rn daughters originating in the LAr and those 
from other background sources.  This analysis focusses on background data collected from August, 2008,  
until just after the argon top-up in November.

\section{Backgrounds in DEAP-1 and pulse-shape discrimination}
This analysis focuses on $\alpha$ radiation from isotopes in the $^{238}$U decay chain.  The dominant source is
$^{222}$Rn and its daughters.  
For this data we supply the PMTs with a lower operating voltage than normal operation to ensure that the 
high energy $\alpha$ particles do not saturate the PMT electronics.
$^{222}$Rn decays quickly 
through $^{218}$Po and $^{214}$Po producing three $\alpha$ particles.  The final $\alpha$ decaying isotope is 
$^{210}$Po (138 days) which is supported by $^{210}$Pb with a much longer half-life (22.3 y).  
The initial amount of $^{222}$Rn decays away quickly ($\sim$3d) 
and this analysis looks closely at the daughters.  The hypothesis is that the short lived daughters
will be distinct from the long-lived $^{210}$Po which will end up concentrated on the surface of the detector walls.  
When a $^{210}$Po on the surface decays the $\alpha$ will either travel away from the wall, through the argon and be
detected, or it will burrow further into the wall and the Po nucleus may move into 
the argon, causing nuclear recoils.


We distinguish different background sources by their \textsl{F$_{prompt}$} ratio,  
defined as: 
\[ F_{prompt} \equiv \frac{PromptPE(150ns)}{TotalPE(9\mu s)} \]
where \textsl{PromptPE} is the number of photo-electrons (PE) between 50 ns before the event to 150 ns after
and \textsl{TotalPE} is the number of PE counted for the entire pulse (9 $\mu$s). 
$n$ and $\alpha$ radiation and nuclear recoils have high F$_{prompt}$, while $\beta$s and $\gamma$s are low 
F$_{prompt}$ events.  We can further distinguish among the high F$_{prompt}$ events by their energy, 
an $\alpha$ particle from the decay of $^{222}$Rn deposits energy in the argon an order of magnitude greater 
than a nuclear recoil.
\begin{figure}[h]
  \begin{minipage}[t]{18pc}
    \vspace{0pt}
    \includegraphics[width=18pc]{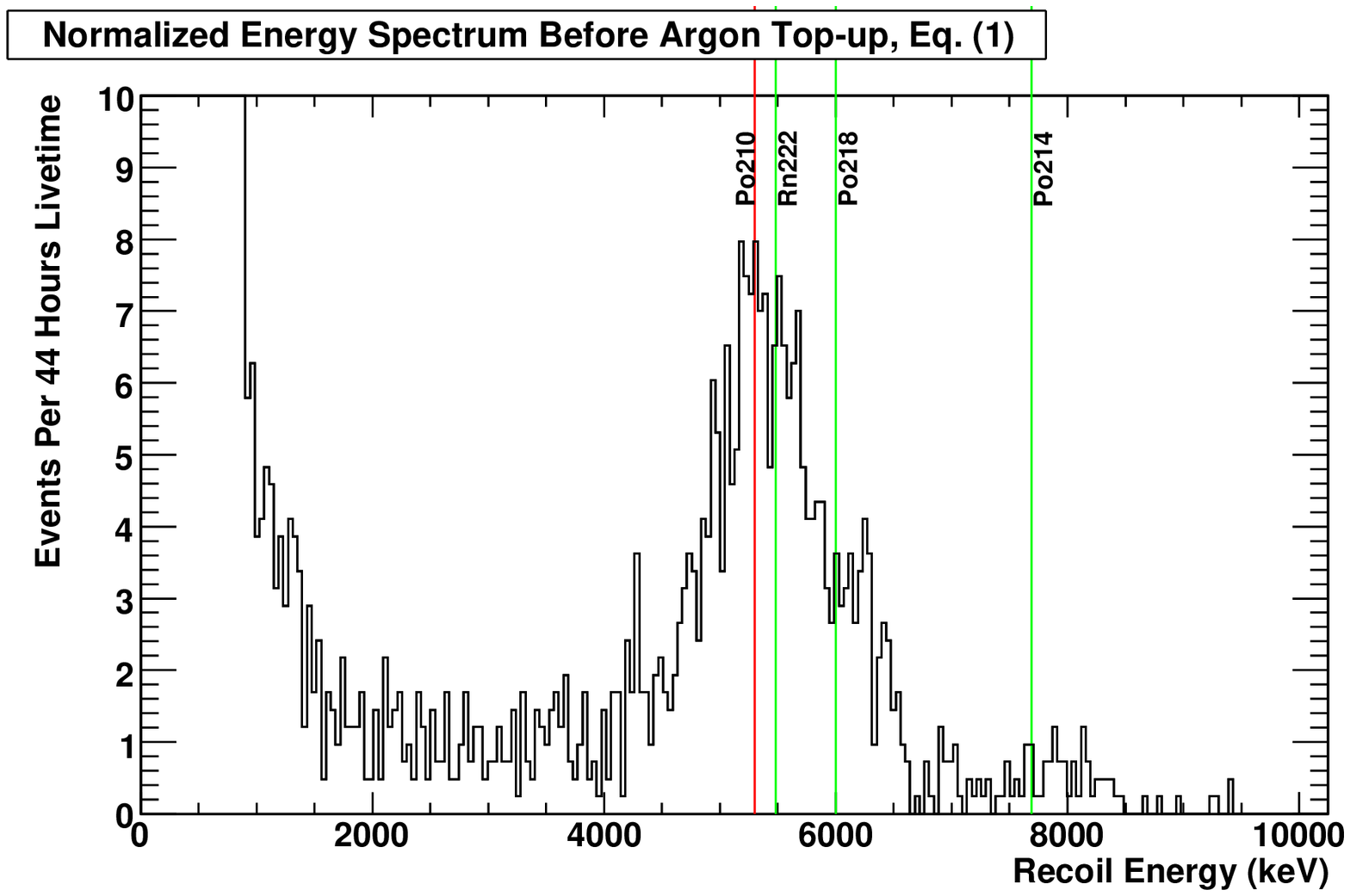}
    \caption{\label{label}Energy spectrum of high F$_{prompt}$ events in the DEAP-1 detector when supplying 
    the PMTs with a lower voltage than normal operation, eq. 1.}
    \label{fig:preboil}
  \end{minipage}\hspace{2pc}%
  \begin{minipage}[t]{18pc}
    \vspace{0pt}
    \includegraphics[width=18pc]{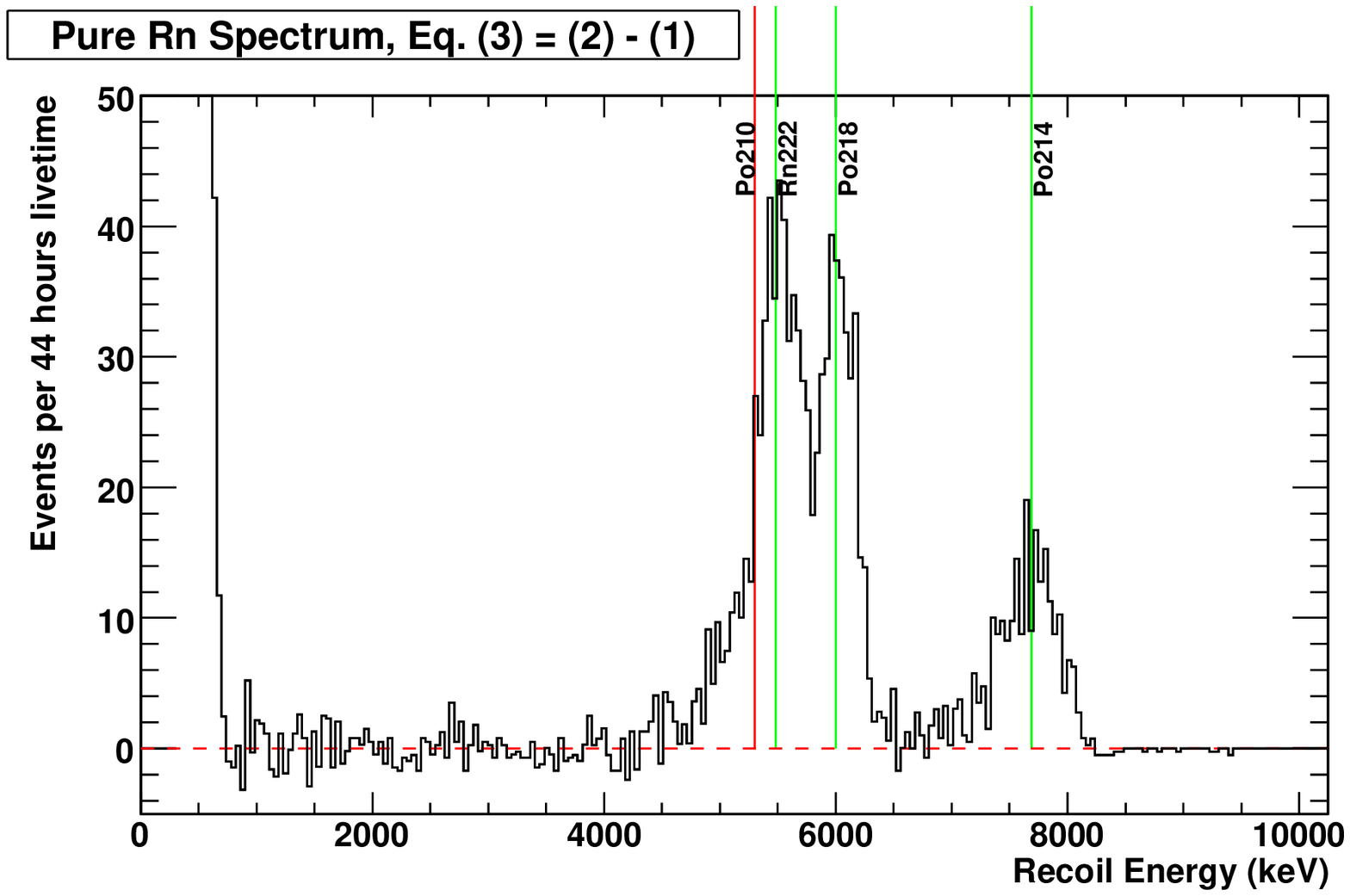}
    \caption{\label{label}Energy spectrum after adding argon to DEAP-1 and subtracting constant background sources, eq. 3.}
    \label{fig:PureRn}
  \end{minipage} 
\end{figure}

\section{Analysis}
Before the November 2008 argon top-up, the energy spectrum of the background can be described by a linear combination
(figure \ref{fig:preboil}):
\begin{equation}
   \frac{dN_{Bg}}{dE}  = \frac{dN^{222}_{Bg}}{dE} +  \frac{dN^{218}_{Bg}}{dE} +  \frac{dN^{214}_{Bg}}{dE} + \frac{dN^{210}_{Bg}}{dE} +  \frac{dN_{other}}{dE}
\end{equation}
Each term represents the number of event counts per energy bin due to each $\alpha$ source, $^{222}$Rn and 
its short lived daughters, $^{210}$Po, and all other backgrounds.  
We are looking at data taken 
within days of the addition, so we treat the long-lived  $^{210}$Po, with a half life of 138 days, as unaffected by 
the top-up.  After the top-up we describe the total backgrounds in the detector from all sources as:
\begin{equation}
     \frac{dN_{Bg}}{dE}  = \frac{dN^{222}_{top+Bg}}{dE} +  \frac{dN^{218}_{top+Bg}}{dE} +  \frac{dN^{214}_{top+Bg}}{dE} +  
     \frac{dN^{210}_{Bg}}{dE} +  \frac{dN_{other}}{dE}
\end{equation}
Since the backgrounds in the detector from sources other than $^{222}$Rn will be equivalent 
in both cases, we take the difference of the two energy spectra.  The remaining events should only be those due to
$^{222}$Rn and its short lived daughters (figure \ref{fig:PureRn}):
\begin{equation}
     \frac{dN_{Bg}}{dE}  = \frac{dN^{222}_{top}}{dE} +  \frac{dN^{218}_{top}}{dE} +  \frac{dN^{214}_{top}}{dE}
\end{equation}
Figure \ref{fig:PureRn} shows that the peaks representing alphas from the decay of $^{222}$Rn and $^{218}$Po 
are now nearly equal in height and that the $^{210}$Po peak shows less contribution. The integral of 
their combined peaks was calculated and compared to the integral of the $^{214}$Po peak. The ratio was determined to be 
3.92 $\pm$ 0.5.

\begin{figure}[ht]
  \begin{minipage}[t]{18pc}
    \includegraphics[width=18pc]{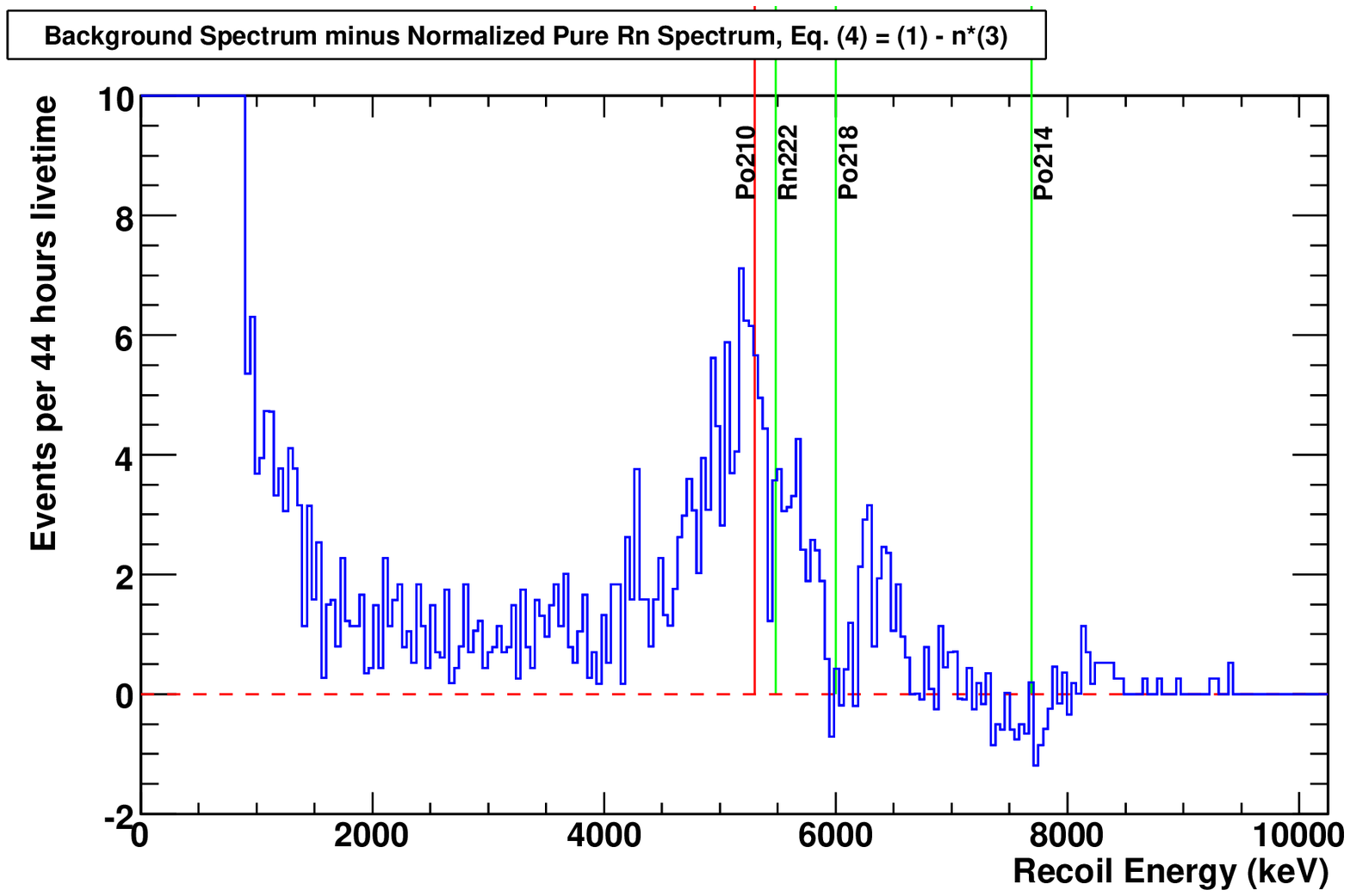}
    \caption{\label{label}Energy spectrum after subtracting pure Rn events.  These are primarily long lived $^{210}$Po 
    decays, eq. 4.}
    \label{fig:diff}
  \end{minipage}
  \hspace{2pc}
  \begin{minipage}[t]{16pc}
    \includegraphics[width=18pc]{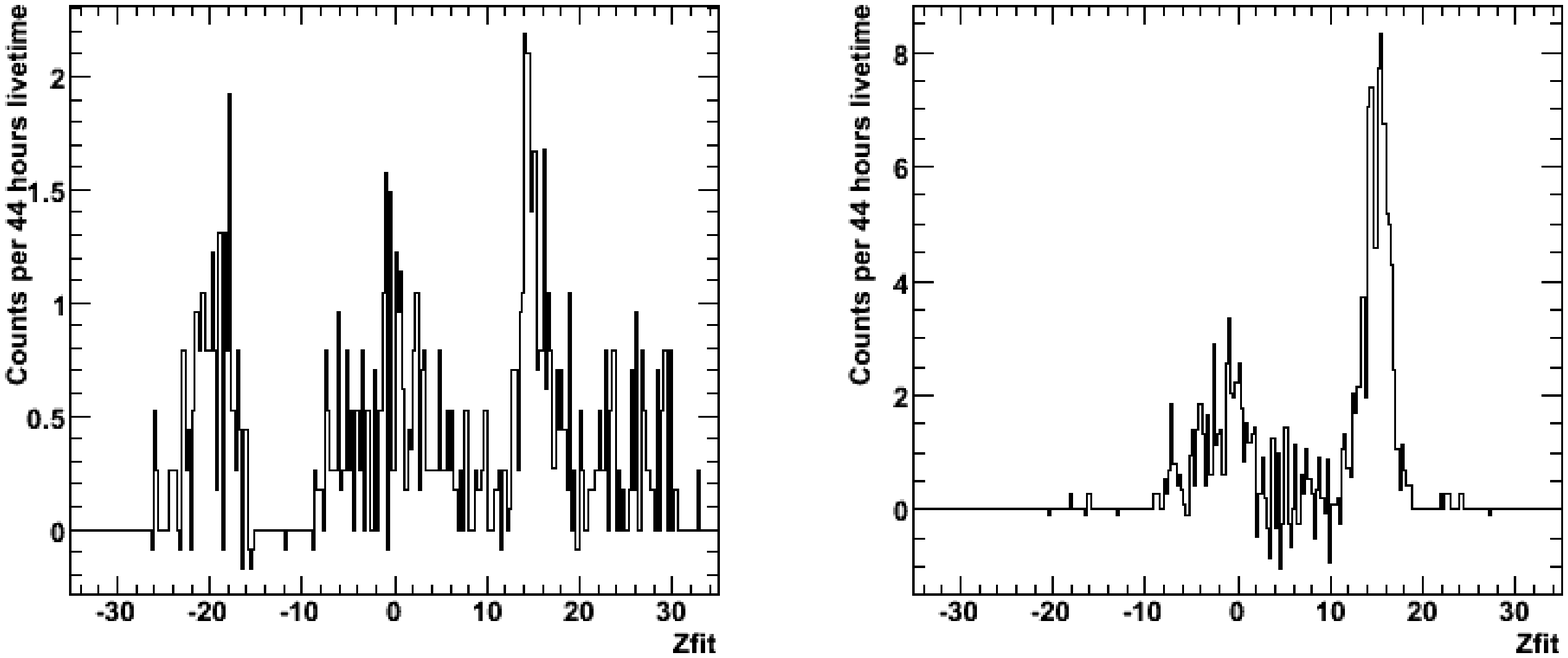}
    \caption{\label{label}Zfit of $^{210}$Po events (right) and lower energy events (1.5--4 MeV, left).
    Backgrounds are concentrated at detector windows.}
    \label{fig:zfitcuts}
  \end{minipage}
\end{figure}

Once an energy spectrum comprised of only those events due to the quickly decaying $\alpha$ emitters from radon is obtained, 
we are able to probe the backgrounds that are less understood: $^{210}$Po and the 'other' high F$_{prompt}$ events.  This 
spectrum was normalized by comparing the integrals of the outlying $^{214}$Po peaks and then subtracted from the 
pre top-up background data described by equation 1.  The result is figure \ref{fig:diff}:
\begin{equation}
     \frac{dN_{diff}}{dE}  = \frac{dN^{210}_{Bg}}{dE} +  \frac{dN^{other}}{dE}
\end{equation}
Figure \ref{fig:diff} has a wide peak
suggesting alpha interactions with variations in light yield.
The over-subtraction visible in figure \ref{fig:diff} is due to small shifts in energy between data collection 
runs and work is ongoing to improve energy calibration.  All events at lower energies are left 
unchanged, and must not be related to the $^{222}$Rn $\alpha$ decays.  Separating the signals due to Rn $\alpha$ 
decays is successful and we see that  $^{210}$Po may be treated as a distinct and constant background source.

To learn about the nature of the remaining events (esp. between 1.5 and 4 MeV) we look to position reconstruction.
In DEAP-1 this is done approximately using the ratio of relative intensities of each PMT, a quantity called Zfit.
The Zfit value of an event is returned in cm and is between $\pm$30 cm.  Due to the difference in efficiency of each PMT
the detector center lies at 7 cm, while the windows are at -7 cm and 15 cm, and the light guides extend an 
additional 15 cm.  The Zfit of the analyzed data  
confirms that the $^{210}$Po accumulates at the detector windows (see figure \ref{fig:zfitcuts}) but also shows 
two distinct regions in which the lower energy events accumulate (those between  1.5 and 4 MeV).  
A significant number of events occur at the interface between the PMTs and the acrylic light guides.  Their source 
requires further investigation.

\section{Conclusion}
This analysis successfully demonstrates that we can distinguish and isolate the signals of background events 
in the detector caused by the decay of short lived $^{222}$Rn and its daughters, as well as the long-living 
$^{210}$Po. In the $^{210}$Po energy spectrum, figure \ref{fig:diff},  a large 
spread of energies and contributions from other events is seen, but, $^{210}$Po decays may be treated as 
a distinct, constant background source.  $^{222}$Rn alphas are distinguishable from other background 
sources and may be reduced by ensuring clean detector components. DEAP-3600 will have the ability to clean and
resurface the inner detector walls, allowing built up $\alpha$ particle contaminants to be removed.

The WIMP region of interest occurs at very low energy in the low-voltage runs, which makes
the identification of these sources important. An estimate of the energy of a
recoiling nucleus is $\approx$750 keV and so cannot be identified in this analysis. 
A similar analysis will follow using nominal voltage background runs in order to improve the data resolution at low
energy.  Starting in March 2009 a radon filter and improved acrylic chamber were installed which reduced 
backgrounds by a factor of 10.

\section*{References}
\bibliography{iopart-num}

\providecommand{\newblock}{}
\begin{thebibliography}{1}
\expandafter\ifx\csname url\endcsname\relax
  \def\url#1{{\tt #1}}\fi
\expandafter\ifx\csname urlprefix\endcsname\relax\def\urlprefix{URL }\fi
\providecommand{\eprint}[2][]{\url{#2}}

\bibitem{KubotaDoke78}
Kubota S, Nakamoto A, Takahashi T, Hamada T, Shibamura E, Miyajima M, Masuda K
  and Doke T 1978 {\em Phys. Rev.\/} B {\bf 17} 2716

\bibitem{HitachiDoke83}
Hitachi A, Takahashi T, Funayama N, Masuda K, Kikuchi J and Doke T 1983 {\em
  Phys. Rev.\/} B {\bf 27} 5279

\bibitem{Kubota82}
Himi S, Takahashi T, Ruan J and Kubota S 1982 {\em Nucl. Instrum. and
  Methods\/} {\bf 203} 153

\bibitem{Lippincott08}
Lippincott W~H, Coakley K, Gassler D, Hime A, Kearns E, McKinsey D~N, Nikkel
  J~A and Stonehill L~C 2008 {\em Phys. Rev.\/} C {\bf 78} 035801

\bibitem{BoulayHime06}
Boulay M~G and Hime A 2006 {\em Astroparticle Physics\/} {\bf 25} 179

\bibitem{MGB09}
Boulay M~G {\em et~al.\/} 2009 {Measurement of the scintillation time spectra
  and pulse-shape discrimination of low-energy $\beta$ and nuclear recoils in
  liquid argon with DEAP-1} (\textit{Preprint}
  \eprint[arXiv]{astro-ph/0904.2930v1})

\end{thebibliography}

\end{document}